# The effects of interstitials clustering on the configurational entropy of bcc solid solutions

Jorge Garcés
Centro Atómico Bariloche, CNEA, 8400 Bariloche, Río Negro, Argentina

This work presents a simple model for describing the interstitials behavior in solid solutions enlarging the current random interstitial atoms paradigm. A general and parameter-free analytical expression to compute the configurational entropy, valid for any tetrahedral or octahedral interstitial solutions and suitable for the treatment of interstitials clustering, is deduced for that purpose. The effect of interstitials clustering on the configurational entropy is shown by applying the methodology to the Nb-H and bcc Zr-H solid solutions. The model for Nb-H presented in this work, based on the existence of H pairs in the α–phase and double pairs in the α'–phase, provides the basis to explain the unsolved controversies in this system. The unusual shape of the partial configurational entropy measured in bcc Zr-H can be accurately described if a small amount of H clusters are included in the solution.

Keywords: entropy, solid solutions, interstitials, Hydrogen, clustering

## 1. Introduction

Despite the significant advances in computational material science in the last years, a quantitative and parameter-free methodology to predict the material properties at finite temperature is still unavailable. However, there are technological requirements to extend the current theoretical methods in order to improve or develop new multicomponent alloys. Consequently, phenomenological models will continue in use unless new theories can be developed in the near future. An illustrative example of the difficulties to develop new and reliable models is related with the calculation of the configurational entropy of mixing in, for example, systems such as amorphous materials or interstitial solutions. Although several models were developed in the last fifty years, it seems that the most suitable lattice gas models for studying configurational properties and phase stability are the Cluster Variation Method (CVM) [1] (with its well known complexity), and the Cluster Site Approximation method (CSA) [2] (simpler at the cost of introducing extra parameters). However, a closed mathematical expression is not directly available by these methods and sometimes the main physical features of the systems, e.g. the right local chemical environment in interstitial solid solutions, could remain unidentified even if a good description of the phase diagram is achieved.

The interstitial solutions belong to a wide field of technological research covering alloys currently of great importance as steels, superalloys, H storage materials, among others. The lack of a simple theory for interstitial alloys limits the advances in this field. The source of the difficulties in modeling these systems arises from the site blocking effects, i.e., there is an effective repulsion or short range order (SRO) between interstitial atoms which may be chemical and/or strain in origin. Moreover, a given site may be simultaneously blocked by more than one interstitial atom (named overlapping or soft blocking) and, therefore, not all lattice sites are equivalent. The main consequence of the SRO is to produce an important decrease in the number of configurations to be computed. Boureau and Tetot [3] showed in their Ti-O study that the number of configurations decreases from $2 \times 10^{19}$ to $2 \times 10^{7}$ if the blocking effect is considered. Consequently, site blocking gives rise to large non-ideal entropy of mixing. This quantity was estimated in the past through Monte Carlo simulations [4] or by analytical methods [5-13] usually limited to low or medium solute concentrations. The clustering among interstitial atoms were not taken into account by all previous methods due to the difficulties of including long-range interactions in the methodology used to compute the number of configurations. Thus, an effect with important consequences on the physical properties of solid solutions remained hidden until now.

Due to the lack of a suitable expression for the configurational entropy of interstitial solutions, some computational codes use the Bragg-Williams approximation and assume ideal mixing in the interstitial sublattice. To overcome the problem of large excess Gibbs energies, a value of β (number of interstitial sites per metal atom) is usually fitted to experimental results. While this procedure is not physically justified, it is the best that can be done at present as quoted recently by Flanagan and Oates [14]. These authors emphasize the need of developing a general and simple expression for the configurational entropy of interstitial solutions based on a model that correctly describes the physics of these systems.

The previous comments show the formidable task that is to properly describe the interstitial solutions and the difficulties to deduce a general expression for these systems; mainly, if the effect of interstitials clustering is included in the formalism. It is clear now that an adequate solution to this long-standing unsolved problem in Material Science will be very difficult to find using traditional methods. The answer to this problem is the main motivation of this work. It will show that the configurational entropy can be computed to a high degree of accuracy, including the effect of interstitials clustering, without the use of intensive computer calculations. This goal is achieved by an interpretation of the configurational entropy problem in terms of probabilities instead of the number of configurations [11]. The developed expression is quite general as it is applicable to tetrahedral (T) or octahedral (O) interstitial solid solutions on any crystal lattice and any interstitial composition. It is suitable to be used as a reference state in Gibbs energy minimization packages and an analytical alternative to the sophisticated Cluster Variation Method and Cluster Site Approximation method. The deduced expression is applied successfully to solve several controversies in the Nb-H and bcc Zr-H systems giving a harmonious picture of the main structural features of these systems.

## 2. Theoretical methods

### 2.1 Probabilistic description of the configurational entropy.

The configurational entropy has been usually calculated through the combinatorial expression $S=k.\ln(W)$, where $W$ is the number of possible configurations with the same energy in the ensemble. All previous models compute the number of configurations based on a lattice gas model under the following assumptions: i) athermal mixture of non-interacting atoms, particles, molecules, or associated chemical species leaving unaffected their internal properties, ii) the equivalence of all N lattice sites, and iii) no superposition between chemical species. It was recently shown that this property can also be deduced by computing probabilities through the inverse problem $S=k.\ln(W)=-k.\ln(P)$ [11]. In this expression $P=1/W$ is the probability of finding a particular configuration of the ensemble formed by all equivalent configurations with the same energy. The probability P is a conditional one and its analytical determination will be usually very difficult. However, if the assumption of independent chemical species in the mixture is used, the conditional probability can be written as a product of independent particles probabilities. Therefore, the configurational entropy can be written as

$$S = -k \ln\left(\prod_i p_i^{n_i}\right) = -k \sum_i n_i \ln p_i \qquad (1)$$

Where $n_i$ and $p_i$ are the numbers and probabilities of each independent species $i$ in the mixture, respectively. Eq. (1) could be a complement to the traditional way of computing the configurational entropy helping to understand and to identify the representative physical features in complicated systems.

### 2.2 Configurational entropy and interstitial clustering.

The probabilistic approach of Eq. (1) is applied in this work to develop a general expression valid for any interstitial solid solution and suitable to describe clustering effects. The independent chemical species in the mixture, i.e. clusters of different sizes, can be identified by doing a right description of the physics of the system using experimental data or *ab initio* molecular dynamic calculations. Whereas the blocking model [5] will be used to describe the repulsive interactions between interstitial atoms, their clustering will model the attractive long-range interactions between them. The blocking model assumes that the occupancy of one site by an interstitial atom excludes the occupancy of $(r-1)$ neighboring vacancy sites. Thus, the set of vacancies is divided in two different species due to SRO: $n_v^f$ free vacancies and $n_v^b = r-1$ blocked vacancies associated with each interstitial atom. The blocked vacancies do not participate in the mixing process and, as a consequence, it could be assumed that a new chemical species of size $r$ is formed. Therefore, the original problem could be reinterpreted as a random mixture of independent chemical species of different sizes: free vacancies of one lattice site size and $n_i$ associated chemical species of sizes $r_i$ formed by the interstitial atoms and their respective blocked vacancies, fulfilling the following relation,

$$\beta N = \sum_i r_i n_i + n_v^f \qquad (2)$$

where $\beta$ is the number of interstitial sites per metal atoms and $\beta N$ the total number of interstitial lattice sites. Thus, the total number of independent species is,

$$\sum_i n_i + n_v^f = \beta N - \sum_i (r_i - 1) n_i = N(\beta - \sum_i (r_i - 1)\theta_i) \qquad (3)$$

Finally, the probability of each chemical species is computed by,

$$p_i = \frac{n_i}{\sum_j n_j} = \frac{n_i}{\beta N - \sum_j n_j^b} \qquad (4)$$

where the indices $i$ and $j$ label all the interstitial clusters and free vacancies in the mixture. The most simple and general expression deducible in the context of Eq. (1), valid for any interstitial solutions including clustering effects, is,

$$\frac{S}{N} = -k\left[\sum_i \theta_i \ln\left(\frac{\theta_i}{\beta - \sum_j (r_j-1)\theta_j}\right) + (\beta - \sum_j r_j \theta_j)\ln\left(\frac{\beta - \sum_j r_j \theta_j}{\beta - \sum_j (r_j-1)\theta_j}\right)\right] \qquad (5)$$

where $\theta_i = n_i/N$ is the composition and $r_i$ the size of each independent chemical species, respectively. These species could be free vacancies, individual interstitials, pairs and double pairs of interstitial atoms, among others.

The composition dependence of each chemical species in the mixture is modeled in this work assuming that the number of clusters is proportional to the number $n_I$ of interstitial atoms as,

$$n_{cluster} = n_I \, f(cluster) \tag{6}$$

where $f(cluster)$ is a function related with the growth of clusters in each solid solution. It will be modeled in this work by a sigmoid function such as,

$$f(cluster) = \frac{A}{1 + e^{-B(\theta - C)}} \tag{7}$$

Thus, the growth of clusters is modeled in this work by a linear function beginning at a critical composition $\theta_c$ related with the parameter C. The parameter A and B describe the amount of clusters in the mixture and the rate of clusters formation, respectively. The parameters are fitted to the partial configurational entropy experimental data and remains invariant for all others calculations. For the Nb-H system, the growth of pairs is also modeled by simpler functions, such as,

i- a linear function: $n_2 = A \, n_I$

ii- a quadratic function: $n_2 = A \, n_I^2$

Eq. (5) takes into account only volumetric effects through the size $r$ of the blocking sphere, composed by $(r-1)$ blocked vacancy sites and the interstitial atom if there is no overlap of chemical species or hard blocking. However, if soft blocking is present in the solid solution, the number of vacancies to be included in the size $r$ depends on the number $n$ of interstitials sharing the same vacancy. In this case, the number of blocked vacancies associated to each interstitial is $(r-1)/n$. Thus, the size $r$ can be deduced analyzing the terminal solubility of the solid solution, usually a compound. The existence or not of soft blocking should be determined and correctly evaluated in that composition. It is important to note that the size $r$ could be the same for all compositions, as in the examples presented in this work, or could change with the temperature and composition if there are different ordered phases in the solutions, as it is observed in the Zr-O system [15]. Eq. (5) shows an easy way to introduce both dependencies in the present blocking model.

*2.2.1 Austenitic Fe-C system.*

The Fe-C in the austenitic phase is a useful system to illustrate the methodology outlined in the Section 2.2. Several expressions were proposed in the past to study this system. All of them were based on a random mixture of isolated C atoms located in the octahedral sites of the fcc structure. A reasonable agreement with experimental information is achieved because the models have adjustable parameters and mainly due to the data to be described are restricted to dilute solutions. Hillert [16] proposed an empirical analytical expression to describe the activity of C atoms for a composition ranging from pure Fe metal to the hypothetical ordered compound Fe$_4$C. This author found that although the expression has no rational ground, the model gave a satisfactory result close to the values used for fitting the experimental data [16].

The partial entropy of mixing for a random mixture of vacancies and one kind of interstitial atom $I$, computed from Eq. (5) is,

$$\overline{S}_I = -kN \left[ \ln\left( \frac{(\beta - r\theta)^r}{\theta(\beta - (r-1)\theta)^{(r-1)}} \right) \right] \tag{8}$$

It is necessary to know the structure of the ending compound of the solution for applying the Eq. (8) to the Fe-C system. Following Hillert, the Fe$_4$C compound with the Pm3m space group is assumed. For this system, $\beta=1$ as there is only one octahedral interstitial site per iron atom. The C atoms occupy the octahedral sites and block all their twelve near neighbors belonging to the interstitial lattice. The structural analysis reveals that the blocking effect has a soft character. Indeed, each blocked site is shared by four C atoms located in the second coordination sphere of the interstitial lattice. Consequently, the number of blocked sites per interstitial atoms is $(r-1)=3$ and the blocking sphere has a size $r=4$. If Eq. (8) is now applied, the expression for the partial configurational entropy is,

$$\frac{S_C}{kN} = \ln\left( \frac{\theta(1-3\theta)^3}{(1-4\theta)^4} \right) = \ln\left( \frac{x(1-4x)^3}{(1-5x)^4} \right) \tag{9}$$

where x is the C molar fraction. The final expression for the C activity computed from Eq. (9) is exactly the same as the one deduced empirically by Hillert, providing the rational basis requested by this author.

*2.2.2 The Ti-O system.*

Another interesting example is the hcp Ti-O solid solution. In the hcp structure, there is only one octahedral interstitial site per metal atom, thus $\beta=1$. The ending compound of the solution is Ti$_2$O. The occupancy of one site excludes two near neighboring sites with a soft blocking character. Consequently, the number of blocking sites per interstitial atom is $(r-1)=1$. The expression deduced by applying Eq. (8) is,

$$\frac{S_C}{kN} = \ln\left( \frac{\theta(1-\theta)}{(1-2\theta)^2} \right) = \ln\left( \frac{x(1-2\theta)}{(1-3x)^2} \right) \tag{10}$$

This expression is the same as the one deduced independently by Moon [7] and Boureau and Campserveux [17]. The general expression (5) deduced in this work provides the theoretical basis to support the use of Moon's expression in hcp systems [16].

*2.3 Ab initio calculations.*

The full-potential linearized augmented plane-wave LAPW method based on density-functional theory as it is implemented in the WIEN2K code [18] is used for the *ab*

*initio* calculations. This code uses the full-potential augmented plane-wave+local orbital (APW+lo) method that makes no shape approximation to the potential or density. The generalized gradient approximation of Perdew, Burke, and Ernzerhof [19] is used for the correlation and exchange. The atomic sphere radii, $R_{MT}$, selected for Nb and H are 2.4 bohr and 1.2 bohr, respectively. Local orbital extensions is included for the semicore states of Nb. The basis set size $R_{MT}K_{max}$ ($R_{MT}$ is the smallest atomic sphere radius inside the cell and $K_{max}$ is a cutoff for the basis function wave vector) is chosen as 4. The charge-density cutoff $G_{max}$ is selected as 22 (in the units system used by the WIEN2K package). The maximum *l* values for partial waves inside the spheres and for the non-muffin-tin matrix elements are $l_{max}=12$ and $l_{max}=6$, respectively. A mesh of 250–300 special *k* points is taken in the whole Brillouin zone. The iteration process is repeated until the calculated total energy, charge, and force components converge to less than $1\times10^{-5}$, $1\times10^{-4}$ and $1\times10^{-3}$ (in the units system used by the code), respectively. A supercell of (2x2x3) bcc unit cell is used to study the relative stability of pairs and one of (3x3x3) for the double pairs. The minimization of forces follows the package and procedures included in the code.

## 3. Discussion and Results

The results of this section will show that an accurate description of the partial configurational entropy allows the identification of the main structural features of the solid solutions. For that purpose, Eq. (5) will be applied to the Nb-H and Zr-H solid solutions in the bcc structure.

*3.1 The Nb-H system*

The H-Nb system has been in the focus of the research for more than three decades. Interesting scientific debates can be found in the literature related with the still unsolved contradictory results regarding the length of the H-H interactions, the nature of the α- and α'-phases and their relations with the ordered structures observed at low temperature and compositions θ=H/Nb>0.75.

Although various models have been proposed in the past, none of them are able to explain all the experimental results in a harmonious frame. This fact shows that they could not identify properly the underlying physics of this system. Indeed, the usually accepted model, i.e., a random distribution of H atoms located in tetrahedral sites, and the physical nature of H-Metal systems are still under current debates. The blocking model approach was recently criticized by Ogawa [10] under the assumption that hydrogen atoms in metal cannot come closer than 2.1Å to each other for all H concentration, an empirical rule known as Switendick's criterium or the ''2-Å rule''[20]. An interesting controversy [21,22] followed this work regarding the approximations and limitations of each model. Although there is experimental evidence of H-H distances lower than 2Å [23-29], the results of this work will show that the criticism of Ogawa is valid in the sense that the interaction among H atoms should be explicitly considered also in the dilute limit. For example, in hcp TM-H solid solutions (TM=Sc, Y, Lu), the behavior of H atoms at medium temperatures is not well described by a random distribution of H atoms over tetrahedral sites. Instead, a random distribution of pair H-TM-H along the c-axis of the cell, where a TM is located in the middle of the pair, is more appropriate [30-32].

Figure 1 shows a comparison between the results of the different models available in the literature for Nb-H [5-12] and the experimental data from Veleckis and Edwards [33]. All these models are based on different methodologies but assuming a random mixture of isolated H atoms. Consequently, the main physical features of this system remain unidentified. Indeed, the model of Boureau is characterized by a hard blocking up to second nearest neighbors. The model of O'Keeffe assumes only one interstitial site per metal atom due to the relaxation around a dissolved interstitial atom. Finally, the model of this work is characterized by a soft blocking of four nearest-neighbors and two second nearest-neighbors sites, as shown in Figure 2a. These models fail to describe the critical composition of the miscibility gap and the experimental solubility limit of H/Nb=1.21±0.04 measured at 750 K [8]. This limit can be described with a blocking sphere of size r=5 (see Figure 2b for details) but is incompatible with the experimental data, as Figure 1 shows.

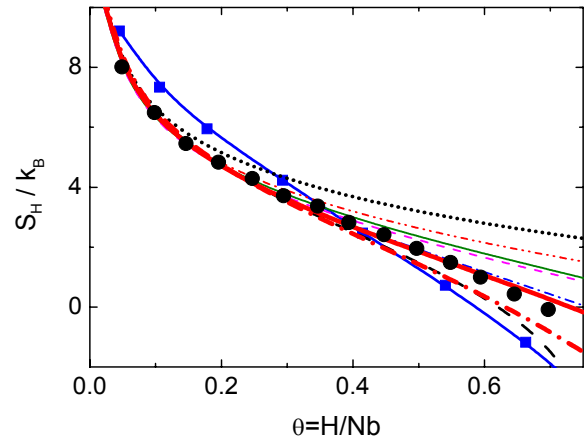

Figure 1. Partial configurational entropy for Nb-H. Comparison among experimental data [33] (full circles) and theoretical models. Thick solid line: this work with blocking size r=4. Thick dash-dot line: this work with blocking size r=5. Solid line plus squares: Monte Carlo calculations [4]. Thin dash-dot line: Boureau´s model [9]. Thick dashed line: Ogawa´s model [10]. Thin dash-dot-dot line: McLellan model [6]. Thin solid line: Speiser and Spretnak model [5]. Thin dash line: Moon model [7]. Short-dot line: Ideal entropy model. The model of O'Keeffe [8] has a similar behavior than Boureau's model. The theoretical results are adjusted to the lowest experimental values. The data for McLellan and Moon models are taken from Ref. [10].

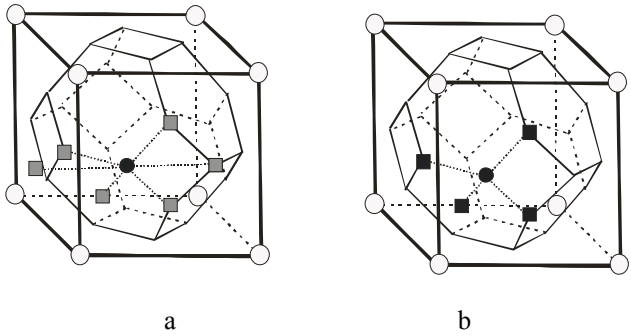

Figure 2. Blocking spheres for the Nb-H system. (a) Soft blocking with size r=4 and (b) hard blocking with size r=5. Empty circles: host lattice atoms. Full circle: H atom. Grey squares: shared vacancies. Black squares: non-shared vacancies. The size r=4 is computed by taking into account that, besides the site occupied by the H atom, each one of the six blocked sites is shared by only two interstitial atoms. The size r=5 arises from a hard blocking of the first four neighbors to the site occupied by the H atom.

It is clear from a comparison among the models shown in Figure 1 that the basic source of controversies in this system arises from the size of the blocking sphere, as shown in Figure 2, and its possible change with the H concentration [4,13]. Other sources of controversies and unsolved problems are the following:

*i-* Length and features of the H-H interaction. Different experimental [34,35] and theoretical [36-42] results, including the calculations of the miscibility gap, the phase diagram and the structures of the ordered hydride phases, suggest that the H-H interaction is characterized by a repulsive interaction extending out to the third or fourth shell of the interstitial lattice and by an elastic interaction energy outside the radius of repulsion. However, these results are in contradiction with the random blocking model assumptions.

*ii-* The nature of the α- and α'- phases is another source of strong controversies as there is experimental evidence of SRO at very dilute H concentrations (< 1 at. %) [43] and in the α'-phase [44,45]. In addition, Shirley and Hall [40] showed that it is impossible to predict simultaneously the α- and α'- and the ordered β- and ε-phases using one set of pairwise interactions of fairly long-range. These authors highlight the importance of the many-body interactions to describe properly the H-H elastic interactions in this system.

*iii-* Regarding the ordered phases, only the phases β and ε have well established structures. The ε-, ξ- and β-phases show no change of domain structure at the phase transition [46]. Seven different phases have been measured in addition to the ε and the *β* phase which have been named as *ι, μ, ν, λ, o, ξ,* and *γ*. They are usually denoted as *λ* phases. According to a structural model for the *λ* phases, sequences of two kinds of *β*-phase microdomains alternate along the *c* direction [47]. However, not all questions concerning this structural model have been solved, e.g. the structure of the microdomain boundaries and the existence of incommensurate modulated phases [48].

The still unsolved problems in H-Nb system, summarized briefly above, raise the question if the H-Nb system is described properly by a random mixture of isolated H atoms. It seems at this point that a different model should be developed to fully describe the physics of the Nb-H system for all H composition.

One common feature of all previous models [3-13] is that they can not include the effect of clustering in the configurational entropy calculation. However, it is possible to compute it easily in the framework of the model presented in this work by considering the different H clusters as new independent chemical species in the mixture. The simplest cluster to be considered to take into account the experimental solubility limit of H/Nb=1.21±0.04, is a pair of H atoms each one with a blocking sphere of size r=5. The expression for the configurational entropy will be computed by applying the present formalism to a mixture of isolated H atoms and pairs. The partial configurational entropy will be computed numerically. The growth of pairs will be modeled by three different functions (see Section 2.2 for details). The values of the parameters, adjusted to the experimental data for the three kinds of pair growth, are: i) linear: A=0.125, ii) cuadratic: A=0.125 and iii) sigmoideal: A=0.10, B=20 and C=0.21. The sigmoideal growth with A=0.10 means that an amount of 18% of the H interstitial atoms are located in pairs. The results of the three models, plotted in Figure 3.a, show that the experimental data can be described better than previous models with any of the three functions proposed for the growth of pairs, although the linear growth shows a remarkable agreement. This result shows that it is possible to solve the contradiction between the blocking size and the experimental solubility limit using a blocking size of r=5 and a mixture of isolated interstitial atoms and pairs.

It is important to point out here that the present methodology can not distinguish among the different kinds of H pairs. All of them are independent chemical species with the same weight. The most stable configuration can be found by a*b initio* calculations. Two possibilities will be considered in this work: i) a pair of two H atoms as close as possible, i.e. the H atoms are fourth-neighbors at a distance of 2.3356 Å, and ii) a pair of two H atoms with a Nb in the middle: H-Nb-H. The H atoms are sixth-neighbors at a distance of 3.6429 Å. *Ab initio* calculations performed with the WIEN2k code give the H-Nb-H pair as most stable with a difference of 0.8mRy/cell between both pair arrangements and 20mRy/cell with respect to two random H atoms. The small energy difference between both pairs could be attributed to a residual H-H interaction in the first configuration. The stability of the second configuration shows the importance of the long range interaction, as suggested in previous experimental and theoretical works [34-40]. Thus, *ab initio* results highlight the relevance of the H-H interaction through the Nb atoms identifying one of the main features of the H behavior in this system.

The good description of the partial configuration entropy achieved by including H pairs in the mixture gives enough confidence in the model to compute the critical composition of the miscibility gap. This calculation is a severe test for any configurational entropy model. In order to compute the maximum of the miscibility gap through the condition $\left(\partial^2 \Delta G / \partial x^2\right)_{xc} = 0$ [13, 48], it is necessary to know the enthalpy of formation of the solid solution. The calculations of the thermodynamical properties in the cluster field approximation by Vaks et al. [49] suggest a considerable temperature dependence and an almost linear behavior of the $\Delta H$ versus $\theta$ in the composition range $0<\theta<0.4$. Consequently, the previous condition in this range is only due to an extreme in the configurational entropy since the non-configurational partial molar entropy of mixing of the hydrogen is assumed, as usually, to be independent of temperature and composition.

The partial configurational entropy derivative for the models plotted in Figure 3a are shown in Figure 3b. While the linear or quadratic growth functions have an extreme in a critical composition $\theta_c>0.4$, the sigmoid function has one located at $\theta_c=0.33$. The last result, very close to the experimental critical composition, highlights the fact that the miscibility gap could be related with an additional structural process, e.g. the clustering of pairs, characterized by a sigmoid growth beginning at a critical composition $\theta c>0$. The present methodology can be applied easily to model the pair clustering if all the pairs in the mixture are assumed to form double or triple pairs and the same parameters obtained from the fitting to experimental data with the sigmoid functions are used. Figure 3.c shows the variation of the extreme of the partial configurational entropy with the amount of double and triple pairs in the mixture. An extreme is located at $\theta_c=0.307$ in remarkably agreement with the accepted experimental value of $\theta_c= 0.31$. These results could explain the controversies regarding the nature of the disordered phases. Whereas the α-phase could be formed by a random mixture of isolated H atoms and pairs, double and triple pairs could characterize the α`-phase.

Although it is possible to obtain a critical temperature in the range of 440 K - 450 K using the data of Vaks et al., the correct procedure for the present model should be to compute the enthalpy of mixing versus composition in function of the independent clusters in the mixture, i.e. individual H atoms, pairs and double or triple pairs. This calculation should be done at finite temperature by means of *ab initio* molecular dynamics, a work that is under current development.

The most stable configuration of the double pair cluster can be identified using *ab initio* calculations. Two arrangements will be selected based on the analysis of the structural evidence related with the ordered phases: i) a square planar arrangement with a Nb in the center and ii) a double pair with two H-Nb-H pairs displaced (½, ½, ½) lattice parameter along the [111] direction. The calculations show that the second configuration is more stable with a difference in energy of 42.7mRy/cell between them.

It is proposed in this work that the most stable double pair configuration could be interpreted as the seed for the ordered phases observed experimentally for H compositions greater than $\theta=0.75$. If this assumption is valid, some evidence should be observed in the phase diagram. Indeed, the phase diagram shows that the β-, ε-and derived λ-phases are only observed below Tc, supporting the existence of a possible relation among these phases and the double pair configuration. In fact, only the ε-, β- and δ-phases have well established structures and all of them can be described by a regular assembly of double pairs. Whereas alternating pure metal and chains-like planes formed by double pairs characterize the ε- and β-phases, as shown in Figure 4, the δ-phases can be interpreted as an arrangement only of chains-like planes.

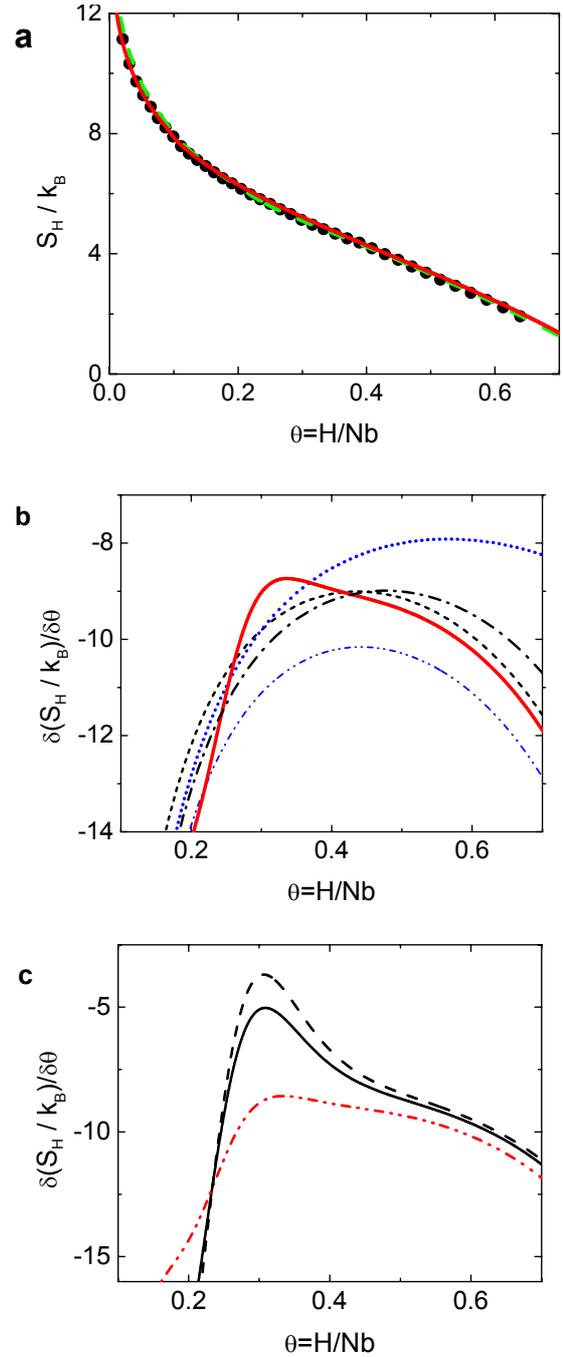

Figure 3. (a) Partial configurational entropy for a random mixture of isolated H atoms and pairs with total blocking size of r=10. Comparison between experimental data [33] (full circles) and theoretical models. Solid line: linear growth. Dash line: sigmoid growth. The experimental error bars are symbolized by the size of the circles. The three models for the growth of pairs are superimposed in the Figure. The theoretical results are adjusted to the lowest experimental value. (b) Location of the critical composition for the following models: i) Solid line: pair clustering with sigmoid growth and blocking size of r=10. ii) Dash-dot line: quadratic growth. iii) Short dash line: linear growth. iv) Dash-dot-dot line: random mixture with r=5. v) Dot line: random mixture with r=4. (c) Change of the critical composition with the amount of pairs clustering. Solid line: 100% of double pairs. Dashed line= 100% of triple pairs. Dash-dot-dot line: 100% of pairs. An extreme is found at xc=0.307 if all the pairs are assumed to form double or triple pairs, in remarkably agreement with the experimental value of xc= 0.31.

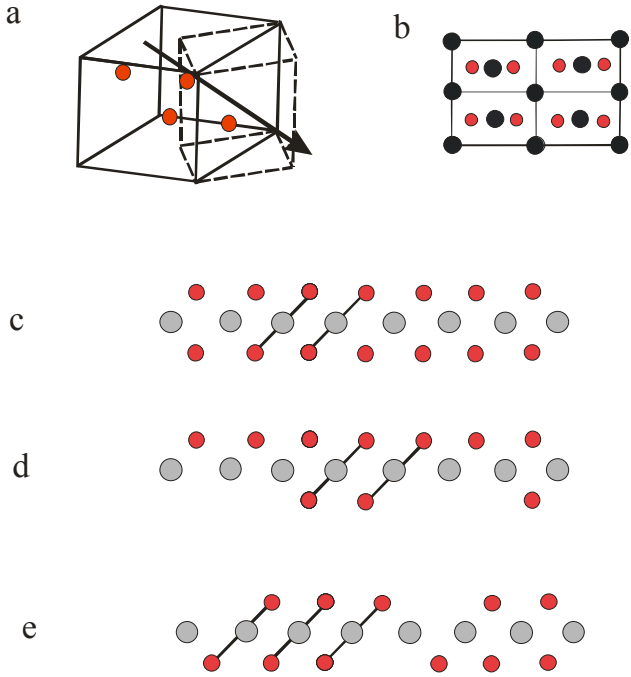

Figure 4. Regular assembly of double pairs in ordered structures. (a) Relation between the bcc cell and the β-phase structure for θ=1. (b) View of the (110) planes along the arrow of Figure 4.a. This is a common view for all ordered structures in Nb-H system. (c) View of the chain along the [$10\bar{1}$] direction of the β-cell (θ=1). (d) The chain for the ε-phase (θ=0.75). (e) Idem for the alternative phase with composition θ=0.75.

The accepted structure for the β-phase with the space group Pnnn (θ=1), can be interpreted as an arrangement of double pairs along the [$10\bar{1}$] direction of the β-cell (equivalent to the [111] direction in the bcc cell), as shown in Figure 4.c. Each Nb atom shares 4 H atoms in a planar arrangement and each H atom shares 4 Nb atoms in a diamond like arrangement. There is no β-structure identified for the composition θ=0.75. The structure for the ε-phase with the space group $P2_12_12_1$ can be built by a regular assembly of double pairs with two vacancy sites on the same side of the chain (50% filled) to adjust the composition at x=0.75, as Figure 4.d shows. An alternative structure, proposed in this work for this composition, is built by displacing two atoms in such a way that all Nb atoms are connected with a H-Nb-H pair. This structure could also be interpreted as an arrangement of triple pairs, as shown in Figure 4.e. It has two different

H sites with a relation 2:1 as was proposed by Hauer et al [49] to explain their experimental results in the ε-phase. Surprisingly, *ab initio* calculations show that the last structure has energy lower than the accepted one for this composition. The difference in energy between both structures is 12Ry/cell. The magnitude of this difference shows the high stability of the triple pair configuration at T=0 K. In this way, the relation between disordered and ordered phases can be established by the formation of double and triple pairs below Tc and composition θc>0.31.

Based on the structural model for the ε- and β-phases, all the λ-phases could be interpreted as the arrangement of the extra H atoms in the vacancies of a supercell, based on one, two or a combination of the structures show in Figure 4, the periodicity of which will depend on the composition. Although none of the λ-phases have accepted structures [50], the present model gives a plausible explanation for their structures based on the existence of double or triple pairs. The possible existence of incommensurate modulated phases or still unobserved superstructures can be easily understood with this simple model. Moreover, the existence of a devil's staircase at very low temperature [51,52] can not be disregarded. In any case, more experimental and theoretical works are needed to fully understand the ordering behavior of these phases.

*3.2 The bcc Zr-H system.*

Zr-H is another very interesting system with some unexplained results and unsolved controversies regarding the behavior of the H atoms in the bcc phase. It is currently assumed that the H atoms are located randomly in the tetrahedral positions [53-57]. However, McQuillan and Wallbank [55] found some anomalies at concentration as low as θ=0.01 in Ti-H and Zr-H systems and criticized the validity of the assumption of complete randomness in these dilute solutions. In addition, there is also experimental evidence that observed precipitates of the γ-phase in the β-phase of Zr as a planar arrangement fulfilling the relation $(00\bar{1})_\beta // (10\bar{1})_\gamma$ between both structures [58-61].

There are several models [5-12] available in the literature for describing the partial configurational entropy measured experimentally by Douglas [53], as shown in Figure 5. All of them were deduced based on a random mixture of H atoms and assumed a solubility limit of θ=1.5. This limit is obtained within the present model by a blocking sphere of size r=4 composed by 4 near-neighbor and 2 next-nearest-neighbor vacancies with soft blocking (See Figure 2a for details). Most of these models fail to give a reasonable description of the experimental data and, mainly, the shape of the curve remains unexplained.

The unusual shape of the partial configuration data and the previously mentioned anomalies raise the question if the physics of this system is well modeled by a random mixture of H atoms.

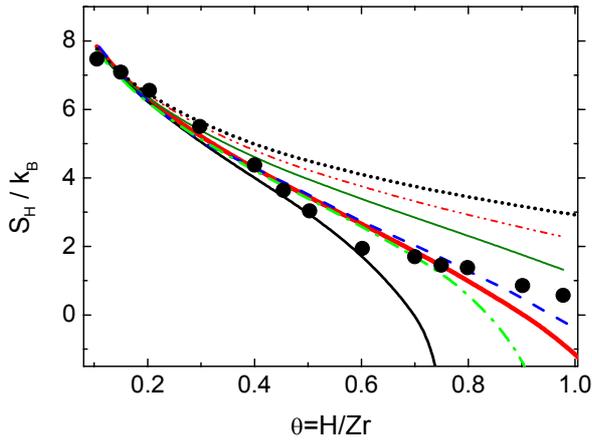

Figure 5. Partial configurational entropy for Zr-H. Comparison among experimental data [52] (full circles) and theoretical models. Thick solid line: this work with blocking size r=4. Thick dashed line: Monte Carlo calculations [4], equivalent to Boureau's model [9]. Thin solid line: Ogawa's model [10]. Thin dash-dot-dot line: McLellan's model [6]. Thin solid line: Speiser and Spretnak's model [5], equivalent to Moon' model [7]. Dash-dot line: O'Keeffe' model [8]. Short-dot line: Ideal entropy model. The theoretical results are adjusted to the lowest experimental values. The shape of the experimental data remains unexplained in Zr-H.

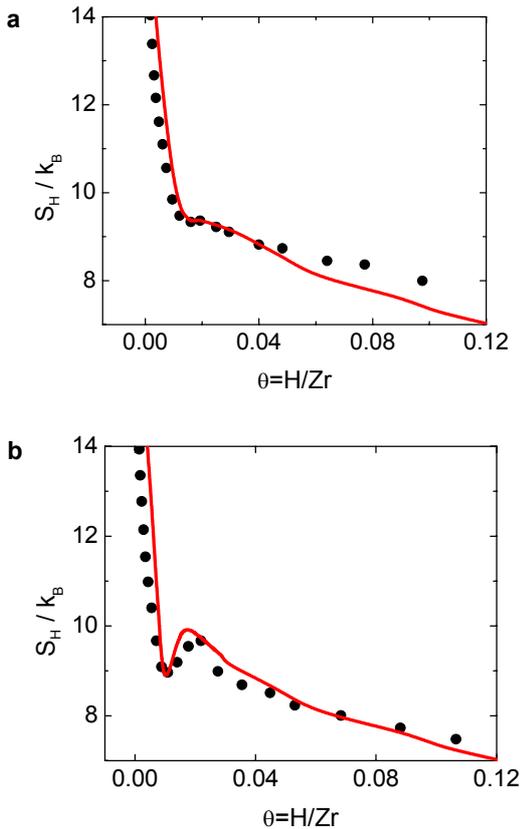

Figure 6. (a) Partial configurational entropy for Zr-H system measured at very dilute H concentrations. Comparison between experimental data [52] (full circles) and the model from Eq. (5) for a mixture of isolated interstitial atoms with blocking size r=4 and pairs. (b) Partial configurational entropy for Ti-H system measured at very dilute H concentrations. Idem Figure 6.a. The theoretical results are adjusted to the lowest experimental value of the relative minimum to compare the shape of both results.

The same methodology applied in Nb-H will be used in this system to look for a phenomenology different to the usually accepted behavior, e.g. the formation of H pairs. If this assumption is valid, a mixture based on isolated H atoms and pairs should explain the anomalies observed by McQuillan and Wallbank.

Figures 6.a and 6.b show that it is possible to describe these anomalies with a sigmoid growth of pairs with parameters A=0.125, B=250, C=0.008 and A=0.125, B=500, C=0.008 for Zr-H and Ti-H, respectively. The result shows that the pair formation begins at very low H concentration and is stronger in Ti-H due to the magnitude of the parameter B, which controls the rate of pair formation.

If this behavior is valid for higher composition, the same parameters used to describe the low composition anomalies should explain the data measured by Douglas. However, the data are well described for a composition lower than $\theta=0.4$ and the unusual shape remains unexplained, as shown in Figure 7.

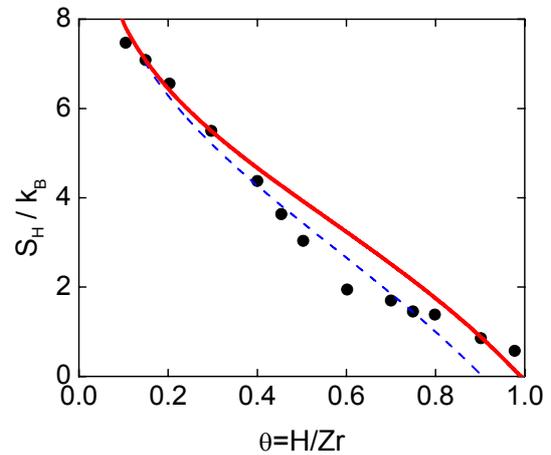

Figure 7. Partial configurational entropy for the Zr-H system. Comparison between experimental data (full circles) and theoretical models. Solid line: the model from Eq. (5) for a mixture of isolated interstitials with blocking size r=4 and pairs. Dashed line: random mixture of isolated atoms with a blocking size of r=4 from Eq. (8).

It is clear from Fig. 7 that an additional phenomenon to the pair formation could exist for compositions beyond $\theta=0.4$. This work proposed that the phenomenon is associated with the double pair formation. The numbers of double pairs are modeled by a sigmoideal growth with parameters A=0.06, B=17 and C=0.52 and the number of pairs beyond the critical composition $\theta=0.4$ will diminish accordingly.

Figure 8.a shows the behavior of the number of pairs and double pairs versus H composition. Figure 8.b shows a comparison between the experimental data and the theoretical result. The agreement is noteworthy if the simplicity of the approach presented here, based on a random mixture of H clusters of different size, is taken into account. This is the first time that it is possible to visualize the effect of H clustering in such a straightforward way.

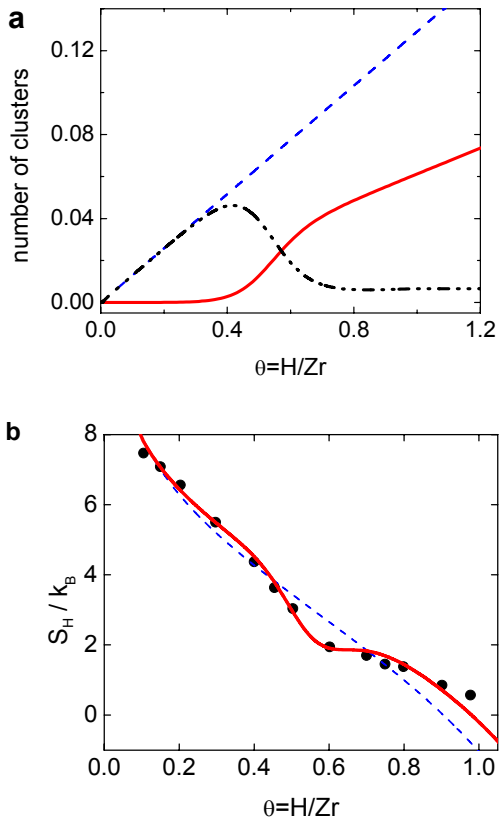

Figure 8. Partial configurational entropy for the Zr-H system. (a) Number of pairs and double pairs versus H composition. Solid line: number of double pairs. Dash-dot-dot line: number of pairs. Dashed line: total number of pairs in the original mixture. (b) Comparison between experimental data (full circles) and theoretical models. Solid line: Eq. (5) for a mixture of isolated interstitials atoms with blocking size r=4, pairs and double pairs. The number of pairs and double pairs for each composition are giving in Figure 6.a. Dashed line: random mixture of isolated atoms with a blocking size of r=4 from Eq. (8). The shape of the theoretical curve is very sensitive to the value of the parameters used to model the growth of double pairs.

**4. Conclusions.**

In summary, a general and parameter-free analytical expression, valid for any interstitial composition on any crystal lattice and suitable for the treatment of interstitials clustering, was deduced using a probabilistic approach to compute the configurational entropy. The basic features characterizing a real system, as the H clustering in bcc solutions can be identified with the help of this simplified expression requiring no intensive use of computers calculations. The expressions deduced in this work, by including the effect of interstitial clustering, open new possibilities in the calculations of phase diagrams and the thermodynamic modeling based on an accurate physical description of the alloys.